\documentclass[twocolumn, tighten]{aastex631}

\newcommand{\RomanNumeralCaps}[1]
    {\MakeUppercase{\romannumeral #1}}
\usepackage{multirow}
\usepackage{amsmath}

\revised{September 18, 2022}
\accepted{September 26, 2022}

\submitjournal{ApJ}

\shorttitle{AGN feedback from X-ray studies}
\shortauthors{Kar Chowdhury et al.}

\graphicspath{{./}{figures/}}

\begin{document}

\title{Cosmological Simulation of Galaxy Groups and Clusters-II: Studying Different Modes of Feedback through X-ray Observations}


\correspondingauthor{Rudrani Kar Chowdhury}
\email{rudranikarchowdhury@gmail.com}
\email{lixindai@hku.hk}

\author{Rudrani Kar Chowdhury}
\affiliation{Department of Physics, Presidency University, Kolkata - 700073, India}
\affiliation{Department of Physics, The University of Hong Kong, Pokfulam Road, Hong Kong}

\author{Suchetana Chatterjee}
\affiliation{School of Astrophysics, Presidency University, Kolkata - 700073, India}
\affiliation{Department of Physics, Presidency University, Kolkata - 700073, India}
\email{suchetana.physics@presiuniv.ac.in}

\author{Ankit Paul}
\affiliation{Department of Physics, Presidency University, Kolkata - 700073, India}

\author{Craig L. Sarazin}
\affiliation{Department of Astronomy, University of Virginia, Charlottesville, VA 22904-4325, USA}
\email{sarazin@virginia.edu }

\author{Jane Lixin Dai}
\affiliation{Department of Physics, The University of Hong Kong, Pokfulam Road, Hong Kong}

\begin{abstract}

The impact of feedback from Active Galactic Nuclei (AGN) on the cosmological evolution of the large scale structure is a long studied problem. However, it is still not well understood how the feedback energy couples to the ambient medium to influence the properties of AGN host galaxies and dark matter halos. In this work we investigate different modes of AGN feedback and their effect on the surrounding medium by probing the diffuse X-ray emission from the hot gas inside galaxy groups and clusters.
For this purpose, we use the cosmological hydrodynamic simulation SIMBA to theoretically calculate the X-ray emission from simulated galaxy clusters/groups with the help of the Astrophysical Plasma Emission Code (APEC). We also perform synthetic observations of these systems with the \textit{Chandra} X-ray telescope using the ray-tracing simulator Model of AXAF Response to X-rays (MARX).
Our results show that in addition to the radiative wind mode of feedback from the AGN, jet and X-ray mode of feedback play significant roles in suppressing the X-ray emission from the diffuse gas in the vicinity of the black hole. Our mock observational maps suggest that the signatures of AGN feedback from the high redshift objects may not be detected with the instrumental resolution of current X-ray telescopes like \textit{Chandra}, but provide promising prospect for detection of these features with potential X-ray missions such as \textit{Lynx}.

\end{abstract}

\section{Introduction}

The existence of super massive black holes (SMBH) at the centres of massive galaxies in the Universe has been widely established in the literature in the past few decades \citep[e.g.,][]{Soltan82, Eckart96, Ghez98, EventHorizon19}. Enormous amount of energy released by a fraction of the SMBH population, in the form of radiative winds, energetic jets, and other mechanisms \citep{Fabian12, K&P15, Harrison18, Roy21} influence several properties of the SMBH host galaxies \citep [e.g.,][]{D&R88, Kormendy93, k&h00, F&M00, Gebhardt00, Graham01, h&r04, Cattaneo09, K&H13, Salviander15, Fiore17, MP18, Schutte19, deNicola19, Marsden20}. Some of the other signatures of AGN feedback involve: reduced star formation rate in host galaxies \citep[e.g.,][]{Vitale13, Costa15, Harrison17}, absence of cooling flows \citep[e.g.,][]{David01, Peterson03} or the change in slope of the $\rm L_{X} - T$ relation in galaxy clusters and groups \citep[e.g.,][]{Andersson09, Maughan12, Molham20}. Despite these series of observational results, the full theoretical understanding of how the energy released from the black hole couples to their surrounding medium, a phenomenon termed as active galactic nuclei (AGN) feedback, is yet to come \citep [e.g.,][]{S&R98, N&R02, Fabian12, Li18, Heinrich21}. 

One of best routes to detect the direct signature of AGN feedback is the X-ray observations of the diffuse gas inside galaxy groups and clusters \citep[e.g.,][]{Wise07, Mc&N07, Mc&N12, Gitti12, Fabian12, Pandge19ApJ, Eckert21}. Energetic outflows from AGN are believed to remove the diffuse X-ray emitting gas from their vicinity, creating regions that are under luminous in X-rays \citep{Gitti10, Laurence11}. This signature has been also observed in the X-ray surface brightness profiles of the gas surrounding the AGN in galaxy groups and clusters \citep{Chatterjee15, Mukherjee18}, but systematic effects need to be characterized for properly exploiting the significance of the results. To understand the role of AGN feedback in the cosmological context a series of subgrid models of SMBH growth and AGN feedback, that are motivated by observational constraints, have been explored in the literature \citep[e.g.,][]{DiMatteo05, DiMatteo08, B&S09, Thacker06, Sijacki07, Sijacki15, Dubois12, Barai14, T&K15, Weinberger17, Dave19, RKC19, Chadayammuri21, Irodotou21, Koudmani21, Habouzit22}.  

Our work is focussed on investigating the phenomenon of AGN feedback in more detail by looking at different modes of feedback from the AGN modelled in cosmological simulation as well as carrying out synthetic observations of these systems. One of the motivations behind the synthetic observation lies in the realistic comparison of the feedback models with actual observational results found in the literature. We begin by modeling diffuse X-ray emission from the simulated galaxy groups and clusters with the help of the Astrophysical Plasma Emission Code \citep[APEC:][]{Smith01}, followed by performing the synthetic observations of the statistical sample of galaxy clusters in the simulation with the {\it Chandra} X-ray telescope using the photon simulator, namely the Model of AXAF Response to X-rays \citep[MARX:][]{Davis12}. 

We follow the mock observation technique of \textit{Chandra} X-ray telescope proposed by \citet[][hereafter R21]{RKC21} using the cosmological simulation by \citet[][SIMBA]{Dave19}. SIMBA is one of the best state-of-the-art high resolution cosmological volume simulations which includes sophisticated modeling of black hole accretion and feedback from the AGN. This simulation takes into account three different modes of AGN feedback, which are wind, jet and for the first time in the literature, X-ray mode of feedback \citep{Choi12} in which the feedback due to X-ray emission from the accretion discs of black holes is included in the simulation. Most of the other simulations consider either the radiative feedback \citep{DiMatteo05, B&S09}, kinetic feedback \citep{Thacker06, Barai14} or a combination thereof \citep{Dubois12, Weinberger17}. In addition to incorporating a detailed modeling of AGN feedback, SIMBA also implements a rigorous black hole accretion process by considering the torque-limited accretion model \citep{H&Q11, Angles-Alcazar13, Angles-Alcazar15, Angles-Alcazar17} along with the Bondi model \citep{BondiHoyle, Bondi52}. Another advantage of using SIMBA is its metal enrichment model that tracks a number of metal elements in the simulation, which are essential for calculating the X-ray flux in galaxy clusters.

In this paper we try to examine the impact of AGN feedback on their surrounding medium by studying diffuse X-ray emission coming from the simulated galaxy groups and clusters using theoretical models as well as synthetic X-ray observations. Apart from this central goal, this paper also aims to study the robustness of the synthetic observation technique proposed by R21. The method was primarily developed in R21 using the cosmological simulation performed by \citet[][hereafter D08]{DiMatteo08}. However, it is important to test the technique with other cosmological simulation to establish its fidelity, which has been carried out in the current work using SIMBA. Results from this study are consistent with R21, showing that the synthetic observation technique is broadly applicable and does not depend on particular simulation parameters. Besides, we find that feedback from the AGN is responsible in evacuating diffuse X-ray emitting gas from their vicinity, consistent with previous studies \citep[e.g.,][]{Mukherjee18, Robson20, RKC19, RKC21}. We also show that radiative wind feedback alone is not sufficient in evacuating hot gas from the vicinity of the AGN. Jet and X-ray mode of feedback plays an important role in this regard, supporting previous studies \citep[e.g.,][]{Choi15}. 

We find that with the resolution limit of the {\it Chandra} telescope, differences in the X-ray surface brightness in the presence and absence of feedback cannot be resolved at few central pixels for the high redshift objects, while a clear difference can be observed for the objects at low redshift. Our study thus makes an avenue to examine the detection possibility of these phenomena in a range of redshifts with current as well as potential future missions like \textit{Lynx} \citep{Lynx17}. The paper is organised as follows. In Section \ref{sec:sim}, we discuss the cosmological simulation used in this work and theoretical modeling of X-ray emission as well as we briefly mention the synthetic observation technique. In Section \ref{sec:results}, we present and discuss the results obtained from this work. Finally, in Section \ref{sec:conclu}, we summarise the work presented in this paper.

\section{Simulation} \label{sec:sim}

As mentioned before we have used one of the most updated cosmological simulation SIMBA \citep{Dave19} that includes hydrodynamics, dark matter dynamics, radiative cooling as well as star formation and their associated feedback along with the accretion and feedback from the black holes for modeling the X-ray emission in groups and clusters. SIMBA uses the Meshless Finite Mass (MFM) technique introduced in GIZMO code \citep{Hopkins15, Hopkins17} as the hydrodynamics solver and Tree - Particle Mesh approach described in GADGET-2 code \citep{Springel05} for the gravity solver. Photoionisation heating and radiative cooling are incorporated in the simulation with the help of GRACKLE-3.1 library \citep{Smith17}. Subgrid $\rm H_{2}$ based model \citep{K&G11} has been followed for calculating the star formation rates. Various metal elements (H, He, C, N, O, Ne, Mg, Si, S, Ca, Fe) are considered to be produced from Type \RomanNumeralCaps{2} and Type \RomanNumeralCaps{1}A supernovae as well as from the Asymptotic Giant Branch (AGB) stars.

SIMBA employs one of the most updated model of black hole accretion and the associated feedback from the AGN. Most of the previous cosmological simulations use spherical Bondi parametrization to model the accretion of the black holes \citep{BondiHoyle, Bondi52} in which black holes are assumed to self regulate their own growth through AGN feedback, after reaching a threshold mass \citep{DiMatteo05, Hopkins06, Sijacki07, Sijacki15, B&S09, Choi12, Steinborn15}. At this stage, surrounding gas of the black holes are heated to a high temperature or expelled, halting further accretion. However, despite the feedback effect, the limiting factor of the black hole growth might depend on the inflow of gas and the rate at which angular momentum is transferred by gravitational torque in galactic disks \citep{Angles-Alcazar17}. Hence it is important to include an updated model of the inflowing gas in the cosmological simulation. 

SIMBA considers the physically motivated angular momentum transport model during the gas accretion onto the black hole from the inner galactic disk \citep{H&Q11, Angles-Alcazar13, Angles-Alcazar15, Angles-Alcazar17}. This torque-limited accretion model is applied to the cold gas $(T < 10^{5} K)$, whereas Bondi accretion is considered for the hot gas $(T>10^{5} K)$ as they are prone to a spherical geometry. Black holes are seeded in the galaxies on-the-fly in SIMBA using friends-of-friends (FOF) algorithm. If a galaxy exceeds a certain critical stellar mass but does not host a black hole, then the star particle nearest to the centre of mass of that galaxy is converted to a seed black hole. Galaxies having threshold mass $\gtrsim 10^{9.5} M_{\odot}$ are populated with the initial black hole of mass $10^{4} h^{-1} M_{\odot}$. 


SIMBA incorporates a novel sub-grid technique for feedback from the black hole that includes three different modes of AGN feedback, i.e, radiative winds, collimated jets and for the first time in cosmological simulation the X-ray mode of feedback. Kinetic feedback is implemented in SIMBA to model the jet mode as well as the effect of radiative feedback in terms of the energetic winds. Outflow from the black holes with high Eddington ratios are modeled based on the observations by \cite{Perna17}. The outflow velocity is parametrised in terms of black hole mass and is given by 
\begin{equation}
v_{wind} = 500 + 500 (\log M_{BH} - 6)/3 ~~\rm km~s^{-1},
\label{eq:wind}
\end{equation}
where $M_{BH}$ is the mass of the black hole in solar masses and $v_{wind}$ is referred as the wind velocity resulted due to the radiative feedback.

For the AGN having Eddington ratio $f_{Edd} \leq 0.2$, jet mode of feedback sets in, in which the velocity is modelled as 
\begin{equation}
v_{jet} = v_{wind} + 7000 \log(0.2/f_{Edd}) ~~\rm km~s^{-1}
\end{equation}
Thus jet feedback is assumed to exert an additional velocity which is capped to a maximum value of 7000 km/sec, achieved at $f_{Edd} \sim 0.02$.

In addition to the feedback in the form of radiative winds and jets, SIMBA also incorporates the feedback due to the photoionisation heating from the X-ray photons coming from the accretion disk of the AGN \citep{Choi12}. Volume heating rate caused by the X-ray photons are calculated following the prescription of \cite{Choi12} and it is distributed to the gas particles that falls within a specific distance to the black hole determined by the accretion kernel. In this mode of feedback, Non-ISM gas is thermally heated to a high temperature while for the ISM gas, half of the energy is applied kinetically and the rest is implemented as heat. We refer the reader to \cite{Dave19} for a comprehensive discussion of the sub-grid model for black hole accretion and feedback employed in SIMBA.

SIMBA is one of the largest cosmological simulations ran with the box size of 100 $h^{-1}$ Mpc that provides extremely high resolution as well. It has also been run with a smaller volume (50 $h^{-1}$Mpc) in which different modes of AGN feedback have been explored. In addition to the scenario in which all the feedback modes mentioned before are included, a no feedback from the AGN is included in the simulation. SIMBA has also been run in the `NoJet' and `NoX' modes. Jet mode of feedback is deactivated in the `NoJet mode' while in the `NoX' mode, both the X-ray and jet mode of feedback are turned off while running the simulation. All the runs have the same initial conditions and use the same cosmology adapted from \cite{Planck16}. Parameters of the simulation runs are shown in Table~\ref{tab:sim_par}.

\begin{table}
	\centering
	\caption{Parameters of different SIMBA runs. $N_{p}$ denotes total number of gas and dark matter particles in the simulation. $m_{gas}$ and $m_{DM}$ denote the initial mass resolution of gas and dark matter particles respectively in different runs of SIMBA. $\epsilon$ represents the gravitational softening length, which in turn indicates maximum resolution of the simulation.}
	\label{tab:sim_par}
	\begin{tabular}{ccccc} 
		\hline
		$L_{box} (h^{-1} Mpc)$ & $N_{p}$ & $m_{gas} (M_{\odot})$ & $m_{DM} (M_{\odot})$ & $\epsilon (h^{-1} kpc)$\\
		\hline
		100 & $\rm 2 \times 1024^{3}$ & $\rm 1.82 \times 10^{7}$ & $\rm 9.6 \times 10^{7}$ & 0.5\\
		50 & $\rm 2 \times 512^{3}$ & $\rm 2.28 \times 10^{6}$ & $\rm 1.2 \times 10^{7}$ & 0.25\\
		\hline
	\end{tabular}
\end{table}

\subsection{Modeling X-ray Emission}
\label{sec:APEC} 

Theoretical X-ray emission is modeled by coupling the simulated gas properties with the Astrophysical Plasma Emission Code (APEC) \citep{Smith01}, using the built in python module ``PYATOMDB". In this work we consider bremsstrahlung emission as the dominant radiative process. We also considered another continuum radiative process known as the e-e bremsstrahlung. In addition to the continuum emission, discrete line emission has also been considered in modeling theoretical X-ray emission. All the information of the emission mechanisms such as collisional and ionisation rates, radiative transition rates, wavelength of transition, energy levels etc. are stored in the atomic database called AtomDB \citep{Smith01, Foster13}. APEC inputs the gas properties into the emission models to produce the emissivity maps. X-ray spectrum is generated by adding all the continuum and line emissivities that is expressed by the following equation 

\begin{equation}
    \epsilon_i= \Lambda_E(T_i,n_{e,i},Z_i)
\end{equation}
where $\Lambda_E$ is generated using APEC by adding continuum and line emissivities with parameters $T_i$, $n_{e,i}$, $Z_i$ that are obtained from the cosmological simulation and they denote temperature, electron number density and metallicity in the $i^{th}$ cell. The integrated X-ray emissivity of the $i^{th}$ cell is obtained by integrating $\epsilon_i$ across the $0.5-2$ keV energy range, i.e,

\begin{equation}
I_i = \int_{0.5-2~keV} \epsilon_i dE
\end{equation}

\begin{figure*}
\begin{center}
\includegraphics[width=16cm]{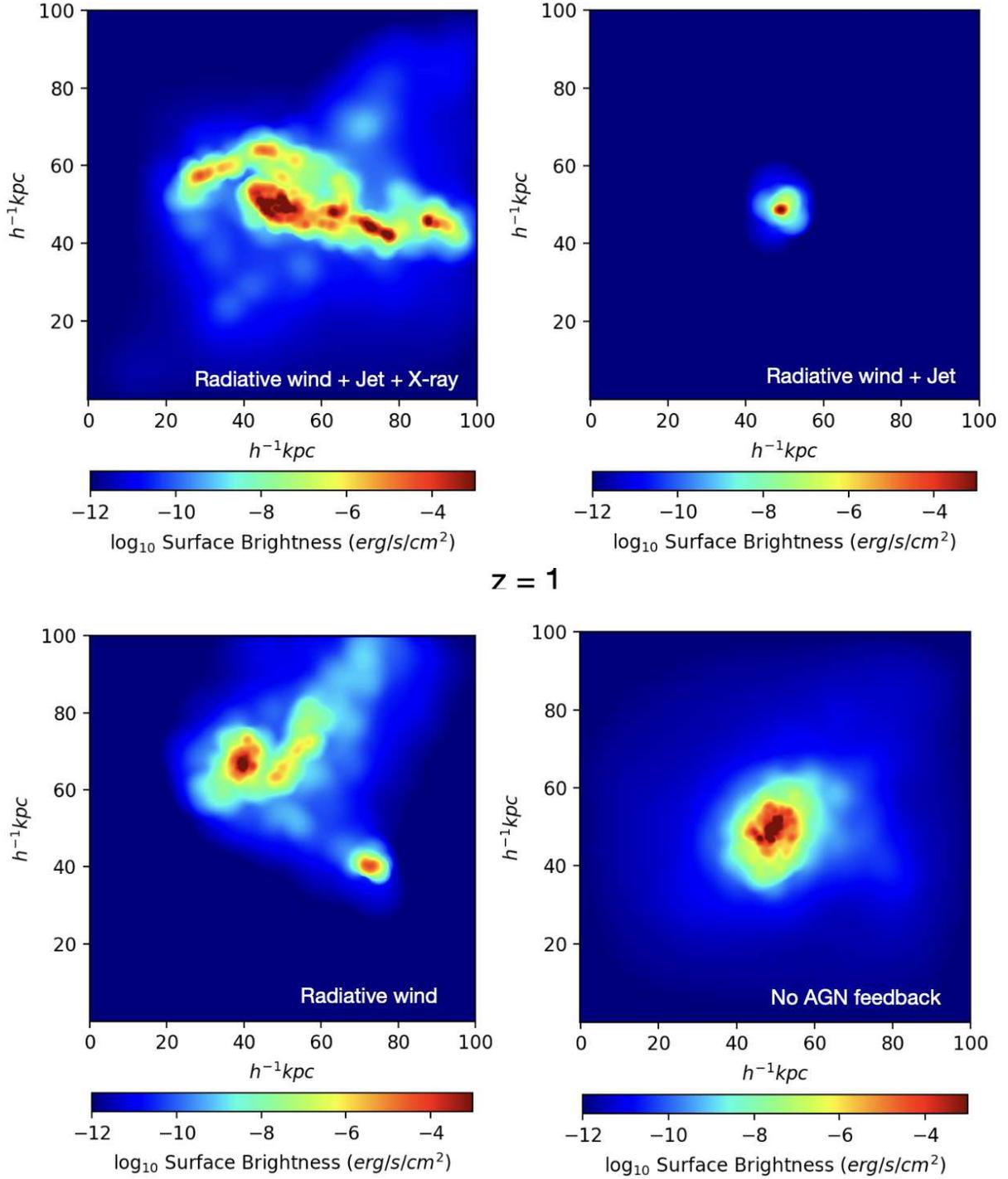}
\caption{X-ray maps of hot gas in the simulations of different feedback modes in the regions of radii 50 $h^{-1}$ kpc having the AGN at the centre at $z=1$. Colourbars show the emitted X-ray surface brightness in $\rm erg/cm^{2}/sec$ calculated using APEC in the soft band (0.5 - 2 keV). This is discussed in Section \ref{sec:APEC}. Top left panel: theoretical X-ray map in the presence of all AGN feedback modes. Top right panel: X-ray map of the hot gas in the vicinity of an AGN in which both the jet and radiative wind mode of feedback are operative but X-ray feedback mode is absent. Bottom left panel: Diffuse X-ray map around the same AGN having only the radiative wind feedback mode. Bottom right panel: theoretical X-ray map of diffuse gas surrounding the same system having no AGN feedback. The statistical results are discussed in Section \ref{sec:results}.}
\label{fig:eg_th_z1}
\end{center}
\end{figure*}

\begin{figure*}
\begin{center}
\includegraphics[width=16cm]{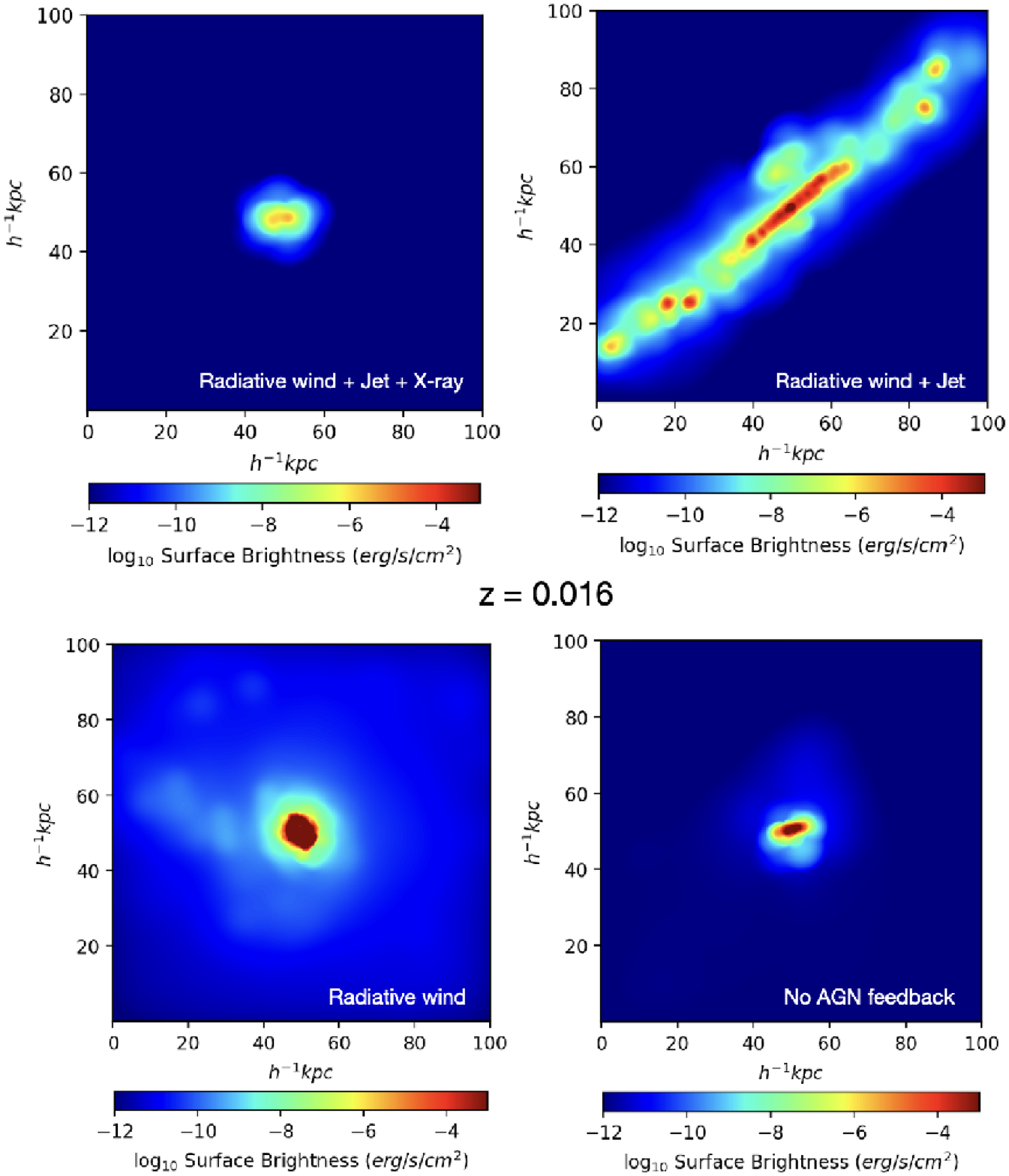}
\caption{X-ray maps of hot gas in the simulations of different feedback modes in the regions of radii 50 $h^{-1}$ kpc having the AGN at the centre at $z=0.016$. Here also colourbars represent theoretical X-ray surface brightness in the soft energy band (0.5-2 keV) in the units of $\rm erg/cm^{2}/sec$ in all the panels. All the four panels have the same layout as in Figure \ref{fig:eg_th_z1}. The linear feature visible in the top right panel, where jet and radiative wind mode of feedback is present, could be the signature of a double sided jet emitted by the AGN at the centre. The stacking analysis is described in Section~\ref{sec:results}.}
\label{fig:eg_th}
\end{center}
\end{figure*}

The X-ray surface brightness thus obtained is then smoothed according to the common B$_2$ smoothing spline \citep{dolag08} given by:

\begin{equation}
 W(r,h) = \frac{\sigma}{h^\nu}
  \left\{
    \begin{array}{ll}
      1 - 6 \big(\frac{r}{h}\big)^{^2} + 6\big(\frac{r}{h}\big)^{^3} ,
      & \,\mbox{$\,\,\,0\,\,\leq \frac{r}{h} < 0.5$} \\
      2(1- \frac{r}{h})^{^3},
      & \,\mbox{$0.5 \leq \frac{r}{h} < 1.0$} \\
      0,
      & \,\mbox{$1.0 \leq \frac{r}{h}$}
    \end{array}
  \right.
\label{eq_mon_kernel}
\end{equation}
where $\nu$ is the dimension of the simulation and its value is chosen to be 3, $h$ is the smoothing length, taken from the simulation at the position of each gas particle and $\sigma$ is the normalisation whose value is fixed at $ 7 \pi /8$. Using this, we generate theoretical X-ray maps of diffuse gas around each SMBH with different modes of feedback within 50 kpc radius region. Example of these are shown in Figure \ref{fig:eg_th_z1} and \ref{fig:eg_th} and discussed in details in Section \ref{sec:results}. It should be mentioned here that the X-ray emission from the AGN itself has not been modeled in this work. As we are focussing on the diffuse emission surrounding the AGN, the signal from the AGN would have the potential to contaminate this \citep{Chatterjee15}. This is discussed again in Section \ref{sec:results}.

\subsection{Synthetic Observation} \label{sec:marx}

In order to do a more realistic comparison with the actual observation and investigate the detection probability of the telescope involved in X-ray observation, we aim to convolve the theoretical X-ray maps with the telescope response functions. For that we have performed synthetic observation with the \textit{Chandra} X-ray telescope using the photon simulator: Model of AXAF Response to X-rays \citep[MARX:][]{Davis12}. We refer the reader to R21 for a detail discussion on MARX and performance of synthetic observations with it. Here we briefly mention the key characteristics of MARX. It is a ray-tracing simulator of the on-board \textit{Chandra} X-ray telescope that minutely tracks all the phases of individual photons from the distant sources. MARX simulates the photons being incident on the high resolution mirror assembly, passed through the diffraction grating and finally detected by the detector. Detailed physics of all the integrated instruments of the telescope are also incorporated in MARX.

We consider diffuse emission from the simulated galaxy clusters as the source while performing the synthetic observation using MARX. All the observations are done in the 0.5 - 2 keV energy band using the ACIS-I detector of \textit{Chandra} for 200 ks exposure time. We have considered the default value of TSTART parameter of MARX, which is 2009.5 for all the synthetic observations performed in this work. Example mock X-ray maps at $z=1$ and $z=0.016$ are shown in Figure \ref{fig:eg_mock_z1} and \ref{fig:eg_mock}, respectively. Details of these figures are discussed in the next section.

\begin{figure*}
\begin{center}
\includegraphics[width=17cm]{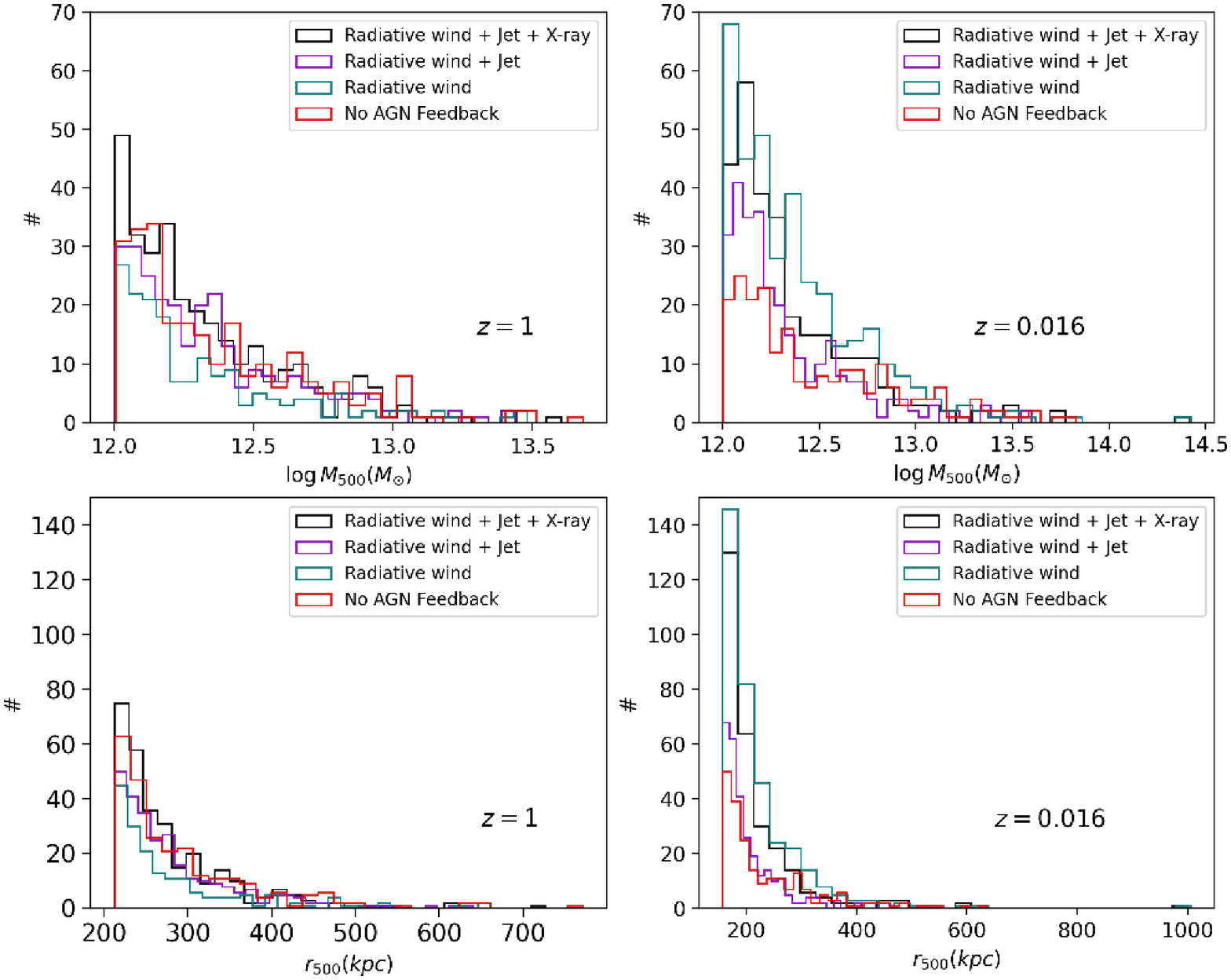}
\caption{Distribution of characteristic mass (top panel) and radius (bottom) of the host dark matter halos of AGN used for stacking in Figure \ref{fig:stack_th}. Left and right panels show the distribution at $z=1$ and $z=0.016$ respectively. In all the panels black, magenta, cyan and red represent the scenario respectively where all, radiative wind and jet, only radiative wind and no AGN feedback modes are present.}
\label{fig:hist}
\end{center}
\end{figure*}

\section{Results and Discussions} \label{sec:results}

\begin{figure*}
\begin{center}
\includegraphics[width=18cm]{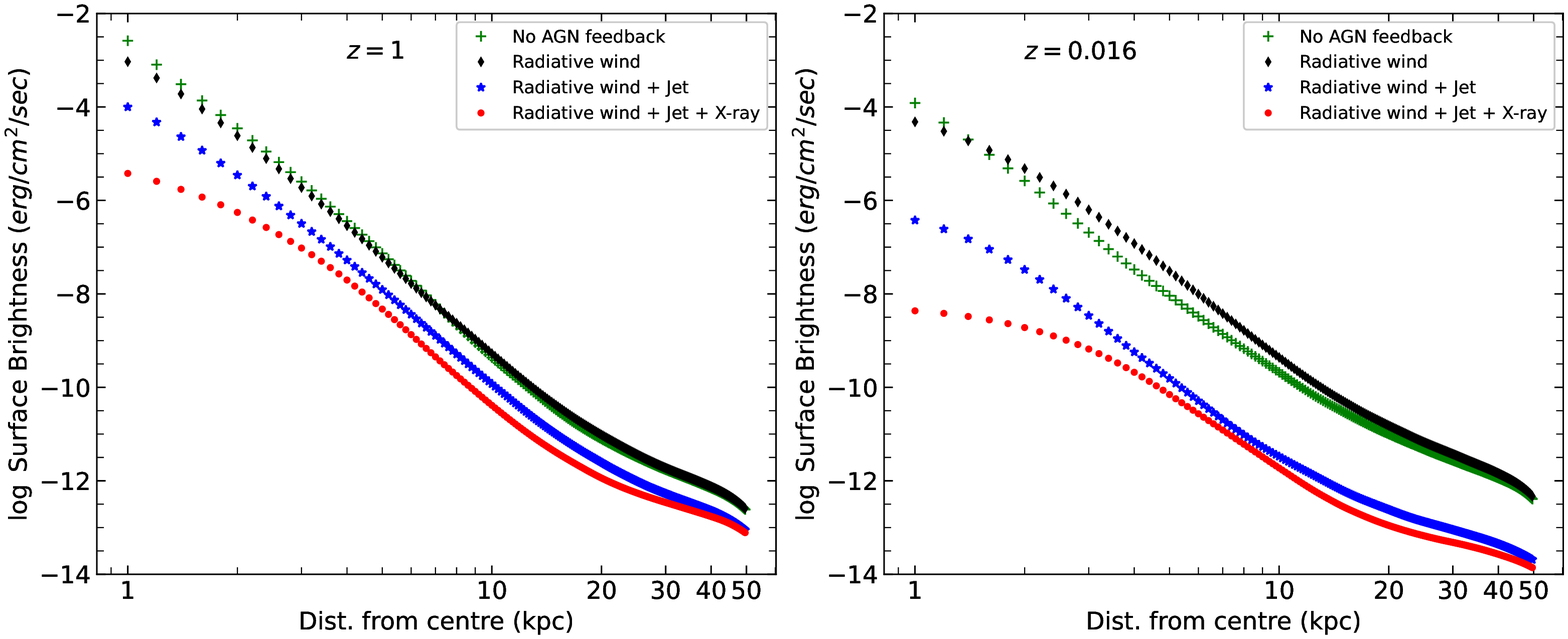}
\caption{Stacked X-ray surface brightness profiles of diffuse gas surrounding the AGN population having different feedback modes. Left panel: radial profile of stacked X-ray emission from the diffuse gas around the AGN with different modes of feedback at z=1 modeled theoretically using APEC. Right panel: same as the left panel except at very low redshift (z = 0.016). Colour schemes are same in both the panels. Green cross represents the scenario when no AGN feedback is present in the systems. Black diamonds stand for the AGNs having only the radiative wind mode of feedback. Blue stars are representative of the cases when both jet and radiative wind mode of feedback are active in the AGN. Red dots show the scenario when all radiative wind, jet and X-ray feedback are present in the simulation. The figure shows that both jet and X-ray feedback are effective at dispersing the surrounding hot gas, but that radiative wind mode alone is not (see Section 3).}
\label{fig:stack_th}
\end{center}
\end{figure*}

In this work, we focus on the impact of the feedback from the AGN on their surrounding medium by investigating the X-ray emission from the diffuse gas inside galaxy groups and clusters. For that, we first theoretically model the diffuse X-ray emission from the simulated groups and clusters at two different redshifts ($z=1$ and $0.016$) in SIMBA as discussed in Section~\ref{sec:APEC}. 

\subsection{Theoretical X-ray Maps}

We have selected central black holes with mass $> 10^{7}h^{-1} M_{\odot}$ inside the host dark matter halo of mass $> 10^{12}h^{-1} M_{\odot}$ from SIMBA for the purpose of this work. Choice of these systems are mainly motivated by the effect of mass resolutions in our simulation. Readers are referred to \cite{RKC19} for a detail discussion of the selection criteria. Systems with different feedback modes at different redshifts have been selected by tracking their host dark matter halos. Figure~\ref{fig:eg_th_z1} and Figure~\ref{fig:eg_th} show examples of  the theoretical X-ray maps of four companion systems having four different modes of AGN feedback at $z = 1$ and $z = 0.016$ respectively. X-ray emission coming from the hot and diffuse gas around the black holes within the regions of radii 50 $\rm h^{-1}$ kpc have been calculated using APEC as discussed in Section~\ref{sec:APEC}. As mentioned earlier, X-ray emission from the AGN has been excluded in our analysis. In both the figures, top left panel shows the X-ray map in the presence of all the feedback modes (radiative wind + jet + X-ray), top right panel represents the scenario when X-ray feedback is absent, bottom left panel is representative of the situation when only radiative wind feedback is active and bottom right panel shows the X-ray maps when no AGN feedback is included in the simulation. Colour code denotes the X-ray surface brightness in units of $\rm erg/cm^{2}/sec$ in both the figures. It can be seen from both the figures that inclusion of all the feedback modes from AGN disperses the surrounding gas, resulting in a lower X-ray emission at the centre compared to the simulation where no feedback from the AGN is included. 

\begin{figure*}
\begin{center}
\includegraphics[width=18cm]{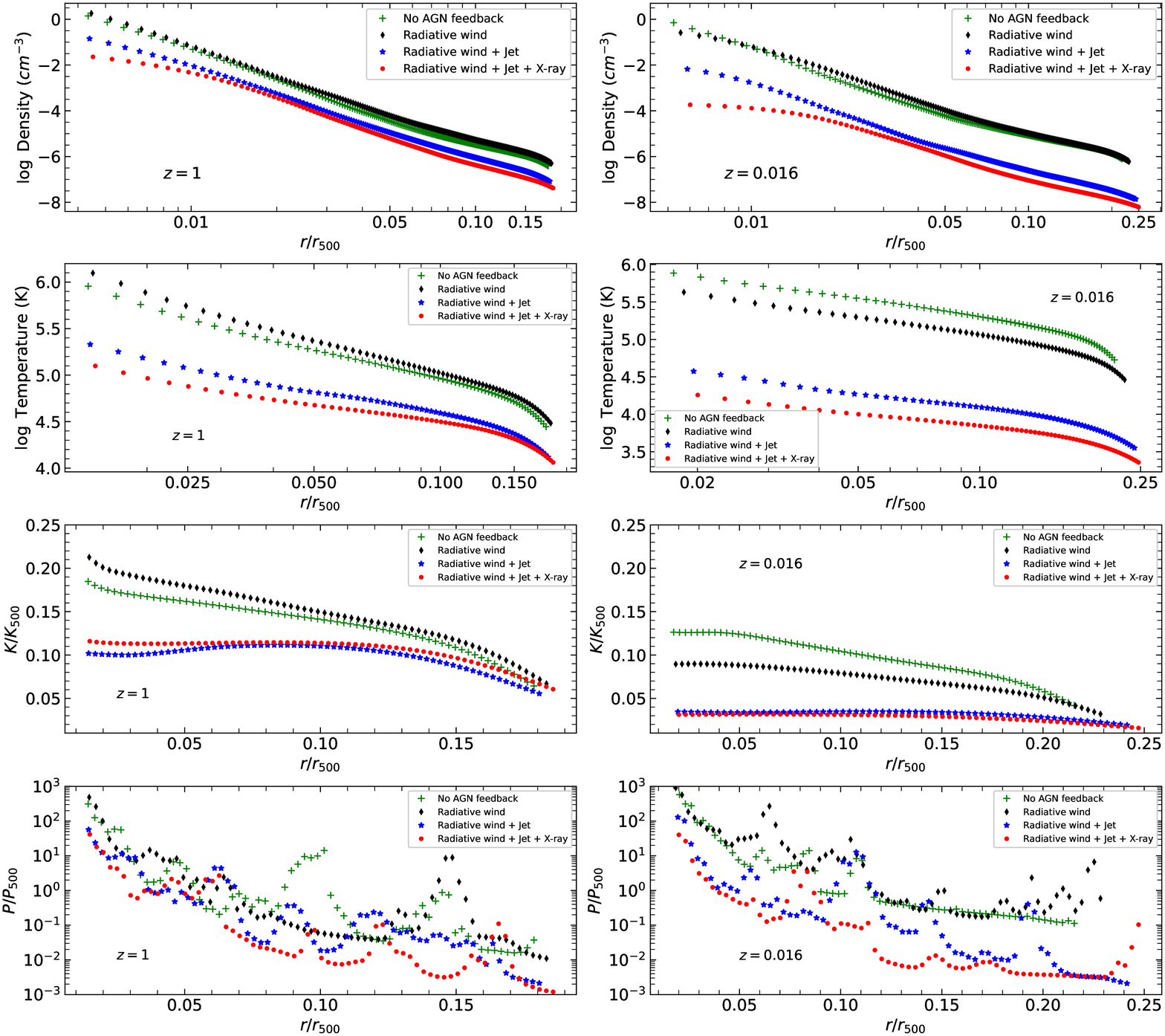}
\caption{Stacked radial profiles of various gas properties. Top panel: Density profiles of gas surrounding the black holes offering different modes of feedback to their ambient medium. Second panel: Stacked radial profiles of temperature in the presence of various feedback modes. Third panel: Entropy profiles at different AGN feedback mode scenario. Bottom panel: Radial pressure profiles in the presence of different AGN feedback modes. Left and right panels represent the profiles at $z=1$ and $z=0.016$ respectively. Colour schemes are same in all the panels and different colour shows different feedback modes similar to Figure \ref{fig:stack_th}.}
\label{fig:pro_all}
\end{center}
\end{figure*}

\subsection{Profiles}

To investigate this phenomenon in more detail, we aim to statistically study the signature of different modes of feedback on their surrounding diffuse X-ray emission at different redshifts. Hence, we select all the systems above the threshold mass mentioned earlier ($M_{BH} > 10^{7} h^{-1} M_{\odot}, M_{halo} > 10^{12} h^{-1} M_{\odot}$) from various feedback mode simulation runs of SIMBA and modeled X-ray emission from the diffuse gas surrounding them within the region of radii 50 $h^{-1}$ kpc. Figure \ref{fig:hist} shows the distribution of characteristic mass ($M_{500}$) and radius ($r_{500}$) of the host halo of the systems that satisfy our selection criteria. We note an abundance of massive halo at lower redshift compared to $z=1$. Figure \ref{fig:stack_th} shows the radial profiles of the stacked X-ray surface brightness obtained theoretically at $z = 1$ (left panel) and $z = 0.016$ (right panel) for different AGN feedback mode scenario. Stacking has been done by taking the average of the surface brightness of all systems.

Both the panels of Figure \ref{fig:stack_th} show that the X-ray surface brightness is significantly lower at the central region when all the feedback modes are included in the simulation at both high and low redshifts. Besides, it can also be seen that although radiative wind mode alone is not sufficient to suppress the X-ray emission at the centre, inclusion of jet and X-ray feedback remove the X-ray emitting gas by large amount from the vicinity of the AGN, thereby reducing the X-ray surface brightness at the centre. 
The minor impact of radiative wind mode feedback on the X-ray surface brightness profile is a consequence of the AGN feedback model incorporated in SIMBA. It relies on a fixed momentum input which results in an insufficient energy injection in the presence of the radiative wind described by Eq \ref{eq:wind}. We refer to \cite{Dave19} for a thorough discussion on this. However, it can be mentioned here that the wind power in SIMBA is weak and could have underestimated the wind feedback \citep{Proga00}. Interestingly, Figure \ref{fig:stack_th} shows an overall suppression of the X-ray surface brightness mainly driven by adding the jet mode, whereas inclusion of X-ray mode in addition to the jet feedback leads to a suppression of the diffuse emission at smaller scales. This is anticipated as X-ray feedback is more dominant in the vicinity of the black hole. We also observe similar effects in the Sunyaev-Zeldovich profiles of similar systems (Chakraborty et al., in preparation).

By comparing left and right panels of Figure \ref{fig:stack_th}, we observe that emitted X-ray flux diminishes at the lower redshift. We explain this as a consequence of the higher gas density inside the high redshift halos that is the dominating factor in calculating X-ray emission in galaxy groups and clusters \citep{Sarazin09, Bohringer10}. In addition to this, presence of all the AGN feedback modes evacuates the gas particles from the vicinity of the black holes, leading to the suppression of emission in the central region. This is also reflected in the radial profile of the gas density in the presence of different feedback modes as discussed next.

Along with the stacked X-ray surface brightness profiles for different feedback modes at two redshifts, we have also studied smoothed density, temperature, entropy and pressure profiles of the same systems, shown respectively from top to bottom panels in Figure \ref{fig:pro_all}.  Left and right panels represent the profiles at $z=1$ and $z=0.016$ respectively. All the panels follow the same colour scheme as Figure \ref{fig:stack_th}. Entropy is defined as $K = k_{B}T/n_{e}^{2/3}$\citep{Voit05}. This is normalized to the value of $K_{500}$ in order to incorporate the mass distribution of galaxy groups and clusters in the entropy profile, which is adopted from \cite{nagai07b}
\begin{equation*}
K_{500} = 1963 \left(\frac{M_{500}}{10^{15}h^{-1}M_{\odot}}\right)^{2/3} E(z) ^{-2/3}\rm keV~cm^{2}
\end{equation*}
where $M_{500}$ is the total mass within the radius $r_{500}$ and $E^{2}(z) = \Omega_{m}(1+z)^{3} + \Omega_{\lambda}$, calculated based on the cosmological parameters used in SIMBA.
Pressure is defined as $P = k_{B}n_{e}T$, which is again normalized to $P_{500}$ following \cite{nagai07b}
\begin{equation*}
P_{500} = 1.45 \times10^{-11} \left(\frac{M_{500}}{10^{15}h^{-1}M_{\odot}}\right)^{2/3} E(z) ^{8/3}\rm erg~cm^{-3}
\end{equation*}
Stacked profiles are obtained by first calculating the profiles of each system normalized to the respective $r_{500}$ value and finally taking the average of profiles of all the systems.

Density profiles in Figure \ref{fig:pro_all} at both redshifts indicate a suppression at the centre in the presence of jet and X-ray mode of feedback. This is in accordance with many previous works using different cosmological simulations where a lower gas density was obtained in the presence of feedback from AGN compared to the no-feedback scenario \citep{Gaspari11, LeBrun14, Correa18, RKC19}. Comparing the density profiles at two redshifts, it can be seen that suppression of gas density at low redshift is more at the centre in the presence of jet (blue stars) and X-ray mode of feedback (red dots) compared to the no feedback (green cross) and radiative wind feedback (black diamonds) cases, supporting our previous claim. This is in accordance with the results obtained using the SIMBA data in previous works \citep{Robson20,Robson21}.
From the temperature profiles (second panel in Figure \ref{fig:pro_all}), it is seen that the temperature of the gas in the presence of jet and X-ray feedback modes, is lower compared to that without feedback. However, it is higher in the radiative mode and as reported by \citet{Dave19}, it seems to be more prominent at lower mass halos. Similar results have been reported by \cite{Robson21}, where they have looked at the redshift evolution of the profiles at different halo mass bins using SIMBA data. This result is in agreement with \cite{RKC19} who use D08 simulation and model for AGN feedback in D08 is equivalent to that of `radiative wind' mode.

We also compare the radial profiles with the observations in order to understand the feasibility of X-ray properties of SIMBA galaxy groups and clusters toward making observational predictions. Density and pressure profiles (top and bottom panel of Figure \ref{fig:pro_all} respectively) show reasonable agreement with radial density \citep{Coston08} and pressure profiles \citep{Arnaud10} of the galaxy cluster sample from the representative \textit{XMM-Newton} cluster structure survey (\textsf{REXCESS}) \citep{bohringer07}. Entropy profiles of the galaxy clusters have also been studied with the \textsf{REXCESS} sample \citep{Pratt10}. \cite{Sun09} studied the entropy profiles of galaxy groups using the \textit{Chandra} archival data. Compared to both of these works, the entropy profiles shown in Figure \ref{fig:pro_all} (third panel) are fairly flat at both redshifts over the spatial scale we are interested in. However, the entropy profiles are in agreement with those obtained by \cite{Robson20} using SIMBA runs for different feedback modes. We note that the observed entropy profiles show a shallower slope at the very central regions. \cite{Pratt10} have discussed that the entropy slopes are shallower at the low mass systems of their sample, because at the lower mass halos, non-gravitational interactions are more dominant. Thus, we understand that the flat nature of the stacked radial entropy profile in Figure \ref{fig:pro_all} is likely to be   a manifestation of the low mass halos present in our sample as seen from the characteristic mass distribution in Fig \ref{fig:hist}.

Moreover, in a recent study \cite{Ghirardini19} looked into the thermodynamical properties of 12 galaxy clusters taken from the XMM Cluster Outskirts Project. Pressure and density profiles of this sample are in agreement with our results. They have also found a significant flattening of the entropy profiles from their sample, consistent with the result of this work. This has been confirmed again by \cite{zhu21} who studied the entropy profiles of 47 galaxy groups and clusters observed with different X-ray telescopes (\textit{Chandra, XMM-Newton, Suzaku}) in order to understand the non-gravitational process in those systems. A fraction of this sample shows a flattened entropy profile around $r_{500}$ which is attributed to the gas clumping effect.

Here we need to mention that different X-ray scaling relations of the galaxy groups and their redshift evolution have been thoroughly studied by \cite{Robson20} and \cite{Robson21} using SIMBA cosmological simulation. Scaling relations obtained using SIMBA are in agreement with the results obtained with different simulations models, such as cosmo-OWLS \citep{LeBrun14}, C-EAGLE \citep{Barnes17}, {\scriptsize BAHAMAS} \citep{McCarthy17} as well as broadly consistent with the results from various observations \citep{Sun09, Vikhlinin09, Pratt10, Eckmiller11, Anderson15,  Lovisari15}. Scaling relations for different AGN feedback modes have also been investigated by \cite{Robson20} and \cite{Robson21}, where they have found that jet mode of feedback plays an important role in evacuating the hot gas from the halos and reducing their temperature. This is in agreement with our findings from this work.

\begin{figure*}
\begin{center}
\includegraphics[width=17cm]{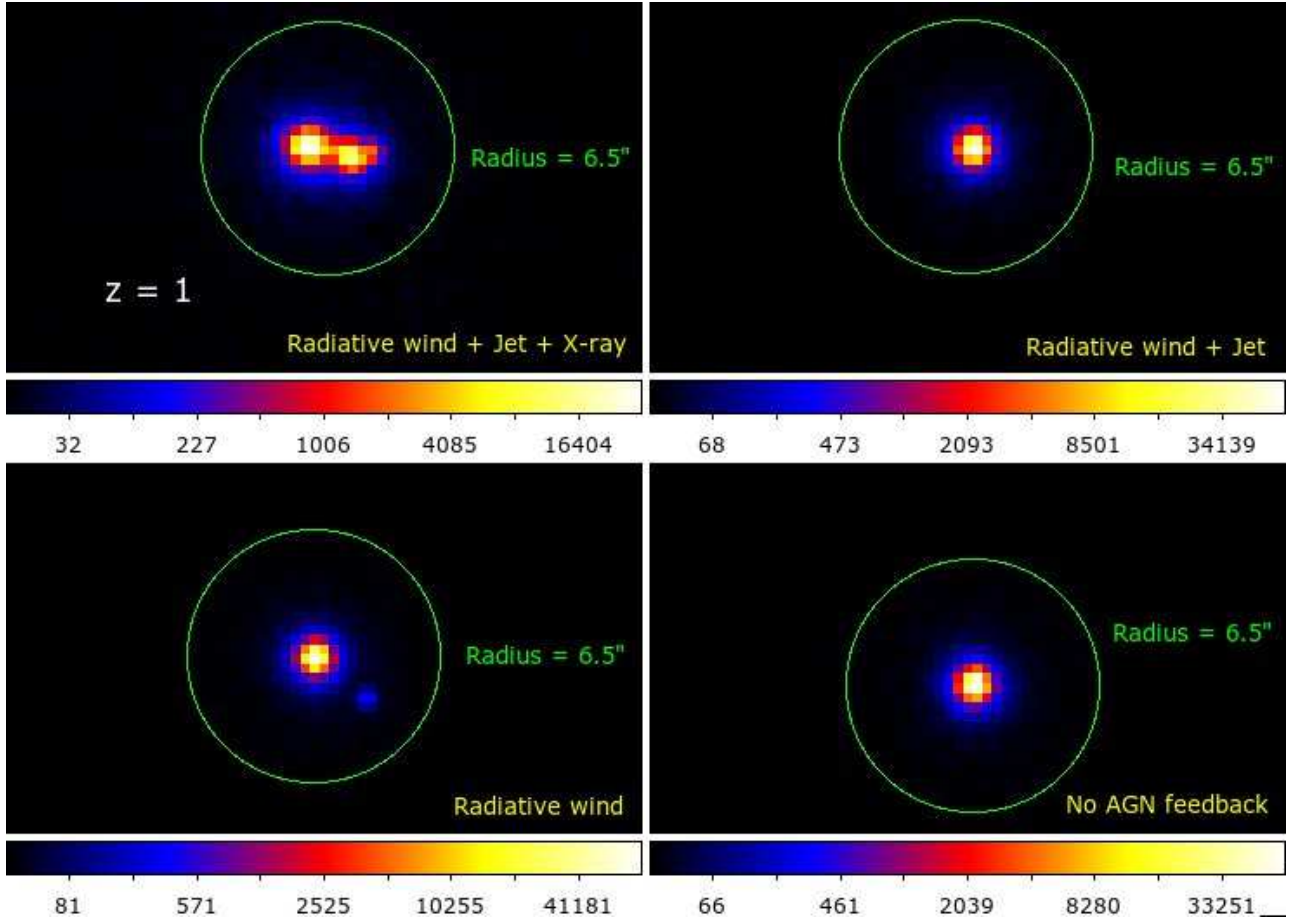}
\caption{Zoom in image of synthetic \textit{Chandra} X-ray maps of the systems shown in Figure~\ref{fig:eg_th_z1}. within the region of radius 50 $h^{-1}$ kpc $\rm (\sim 6.5'')$ at $z=1$. In all the panels colourbars represent the photon number counts per pixel, where the pixel size is $\sim 0.5''$. Top left panel: mock X-ray map of the diffuse gas around an AGN having all the feedback modes. Top right panel: synthetic X-ray map of hot gas in the vicinity of an AGN that provides radiative wind as well as jet feedback to its ambient medium. Bottom left panel: diffuse X-ray map of synthetic observation surrounding an AGN in which only radiative wind mode of feedback is present but jet and X-ray feedback are absent. Bottom right panel: mock X-ray map surrounding a black hole having no feedback mode. See Section \ref{sec:3.3}. for discussions.}
\label{fig:eg_mock_z1}
\end{center}
\end{figure*}

\subsection{Mock X-ray Observations}\label{sec:3.3}
\begin{figure*}
\begin{center}
\includegraphics[width=17cm]{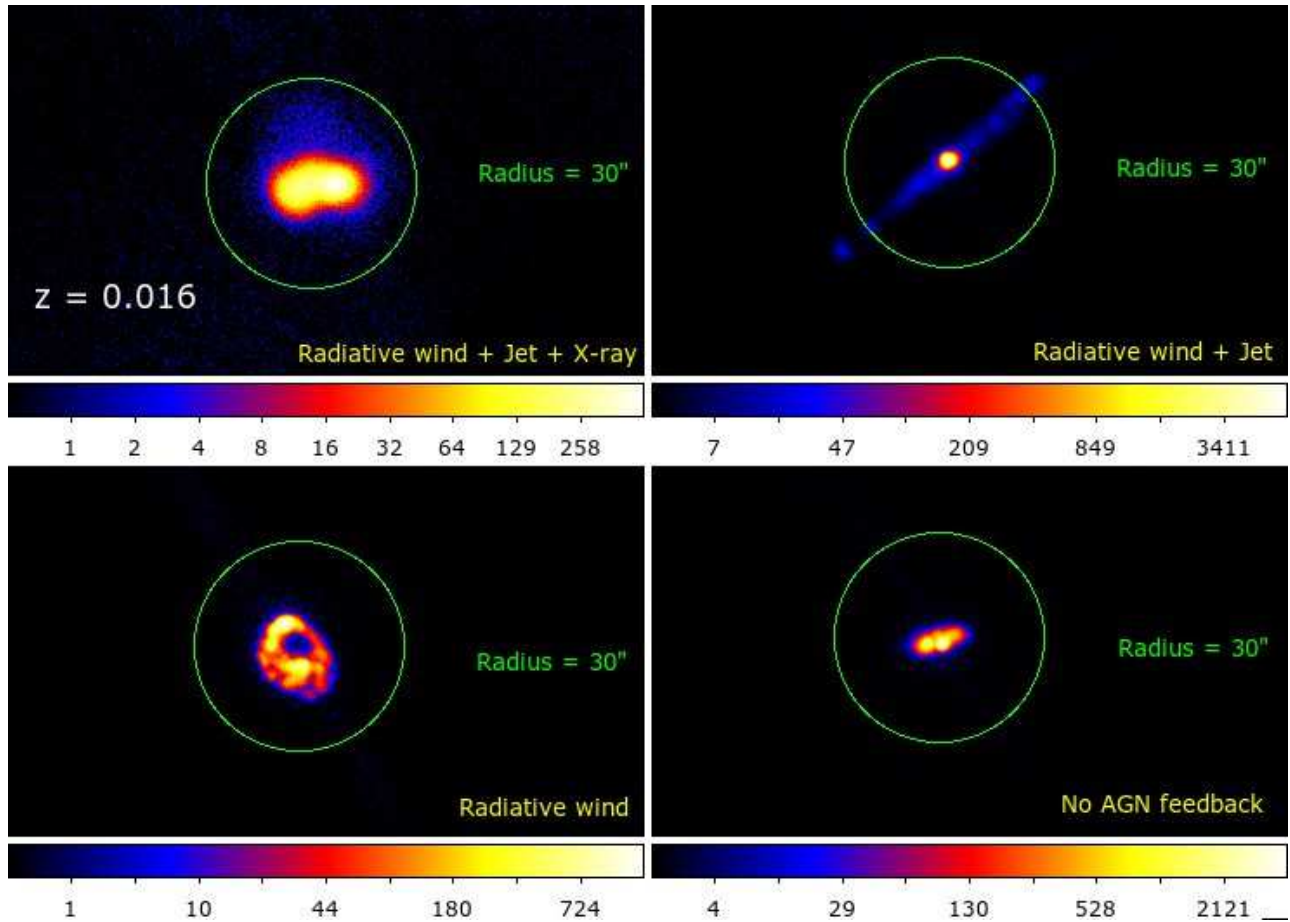}
\caption{Zoom in image of synthetic \textit{Chandra} maps corresponding to the low redshift systems ($z = 0.016$) shown in Figure~\ref{fig:eg_th}. within 50 $h^{-1}$ kpc radius region $(\sim 2.5')$.  Colourbars represent the photon number counts per pixel. The different panels have the same feedback parameters as Figure \ref{fig:eg_mock_z1}.}
\label{fig:eg_mock}
\end{center}
\end{figure*}

After modeling theoretical X-ray emission inside simulated galaxy groups and clusters, we aim to perform synthetic \textit{Chandra} observations of the systems with the help of MARX as described in Section~\ref{sec:marx}. Mock X-ray maps of the systems corresponding to Figure \ref{fig:eg_th_z1} and Figure \ref{fig:eg_th} are shown in Figure \ref{fig:eg_mock_z1} and Figure \ref{fig:eg_mock} respectively. Synthetic \textit{Chandra} observations have been performed for each systems for 200 ksec exposure time in the soft energy band (0.5 - 2 keV) within a region of radius 50 $h^{-1}$ kpc that corresponds to an angular scale of $\rm 6.5''$ and $\rm 2.5'$ at $z = 1$ and $z = 0.016$ respectively. Figure~\ref{fig:eg_mock_z1} and Figure \ref{fig:eg_mock} show the zoom in maps of diffuse gas surrounding the AGNs having different feedback modes. In both the figures, top left panel shows the X-ray maps in presence of all modes of AGN feedback, top right panel represents the scenario when all but X-ray feedback is present, bottom left panel illustrates the X-ray map in which only the radiative wind feedback is present and bottom right panel is representative of the scenario when no feedback is provided from the central black hole. Colourbars represent the photon number counts per pixel in both of these figures where the pixel size is $\sim 0.5''$. No distinctly visible signature of AGN feedback can be identified from the X-ray maps of high redshift objects (Figure \ref{fig:eg_mock_z1}). We explain this as a consequence of limited angular resolution of the \textit{Chandra} telescope, which is $\sim 0.5''$ (see R21). This is in agreement with the findings of R21 (Fig.1 of that paper). However, it is noted that the signature of AGN feedback of the low redshift objects can be resolved from their synthetic X-ray maps (Figure \ref{fig:eg_mock}).

\begin{figure*}
\begin{center}
\includegraphics[width=18cm]{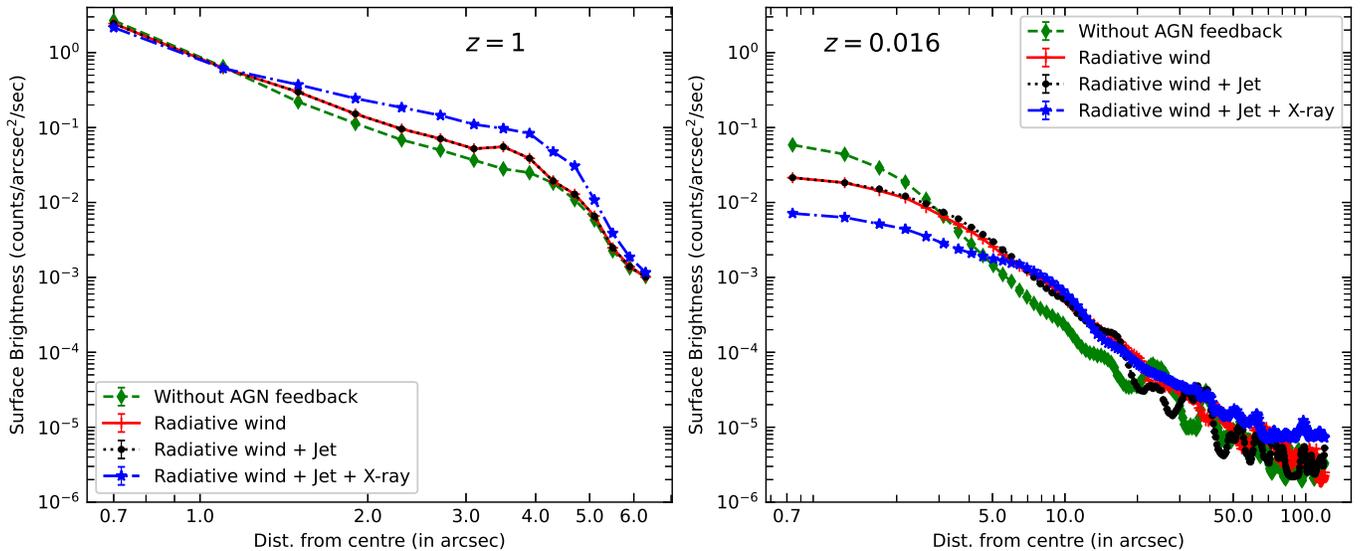}
\caption{Stacked synthetic X-ray surface brightness profiles for different AGN feedback scenario. Left panel: stacked radial profile of X-ray surface brightness obtained from mock observation of diffuse gas surrounding the AGNs having different modes of AGN feedback at z=1. Right panel: same as left panel but at very low redshift (z = 0.016). Colour schemes are same in both the panels. Green diamonds represent the scenario when no AGN feedback is present in the systems. Red lines show the systems having only the radiative wind mode of feedback. Black dots stand for the AGN in which both the radiative wind and jet mode of feedback is active but X-ray feedback is absent. Finally, blue stars are representative of the AGN having all the three modes of feedback. We notice that different modes of AGN feedback are difficult to disentangle at the very central region of the high redshift objects. However, in the low redshift objects, we see a significant decrement of the X-ray surface brightness at the central region when all the feedback modes are included. See Section \ref{sec:3.3} for details.}
\label{fig:stack_mock}
\end{center}
\end{figure*}

We statistically study the imprint of AGN feedback on their surrounding medium to understand the importance of different modes of feedback from the AGN along with their detection possibility at different redshifts. Hence, we perform synthetic X-ray observations of all the systems used in Figure \ref{fig:stack_th} for 200 ksec exposure time using \textit{Chandra} telescope in the soft energy (0.5 - 2 keV) band. Then we stack the total X-ray surface brightness obtained from synthetic observations of all the systems and construct the radial profiles accordingly. Stacked X-ray surface brightness profile obtained from synthetic observations of various AGN feedback scenario at high ($z = 1$) and low ($z = 0.016$) redshifts are shown in Figure \ref{fig:stack_mock}. Same colour schemes are maintained in both the panels. Errorbars represent the standard errors.

No significant difference in the X-ray surface brightness for different AGN feedback modes is seen within few central pixels of the stacked radial profiles of the high redshift objects (left panel). However, a notable excess of the X-ray emission is observed at the outer radius in the presence of all the feedback modes (blue stars). Similar result has been obtained by R21 from the synthetic \textit{Chandra} observation of the simulated galaxy groups and clusters at $z=1$ using the D08 simulation. This indicates that the synthetic observation method results in the same conclusion using two different cosmological simulations. Also, the results are in accordance with the findings of \cite{Mukherjee18} who showed the reduction of X-ray emission from the hot gas inside active galaxies as well as an excess emission in the outer radius detected from deep X-ray observations. We explain the excess X-ray emission at the larger radii to be coming from the accumulation of the hot gas displaced from the centre due to feedback effects. This excess emission could be a viable signature of AGN feedback of the high redshift objects where the feedback features cannot be detected at the centre due to limited angular resolution of the telescope involved. 

From the stacked radial profiles of the low redshift objects including different feedback modes in the right panel, a clear difference in the X-ray surface brightness is visible between various feedback modes within the central pixels. X-ray surface brightness is suppressed in the vicinity of the AGN in the presence of all three feedback modes compared to the scenario when the central black hole provides no feedback to the surroundings. It is also obvious from the plot that inclusion of X-ray mode of feedback along with the radiative wind and jet mode decreases the X-ray emission significantly surrounding the black hole. An excess emission is observed also in these low redshift objects at the larger radii when all feedback modes are active compared to the scenario when no feedback is offered by the central black holes (green diamonds). From Figure \ref{fig:stack_mock} we observe that it is possible to resolve the AGN feedback signatures with \textit{Chandra} from the low redshift objects while for the high redshift sources, excess emission in the AGN host systems at the outer radii could be a plausible signature for detecting feedback effects.

\begin{figure*}
\begin{center}
\includegraphics[width=17cm]{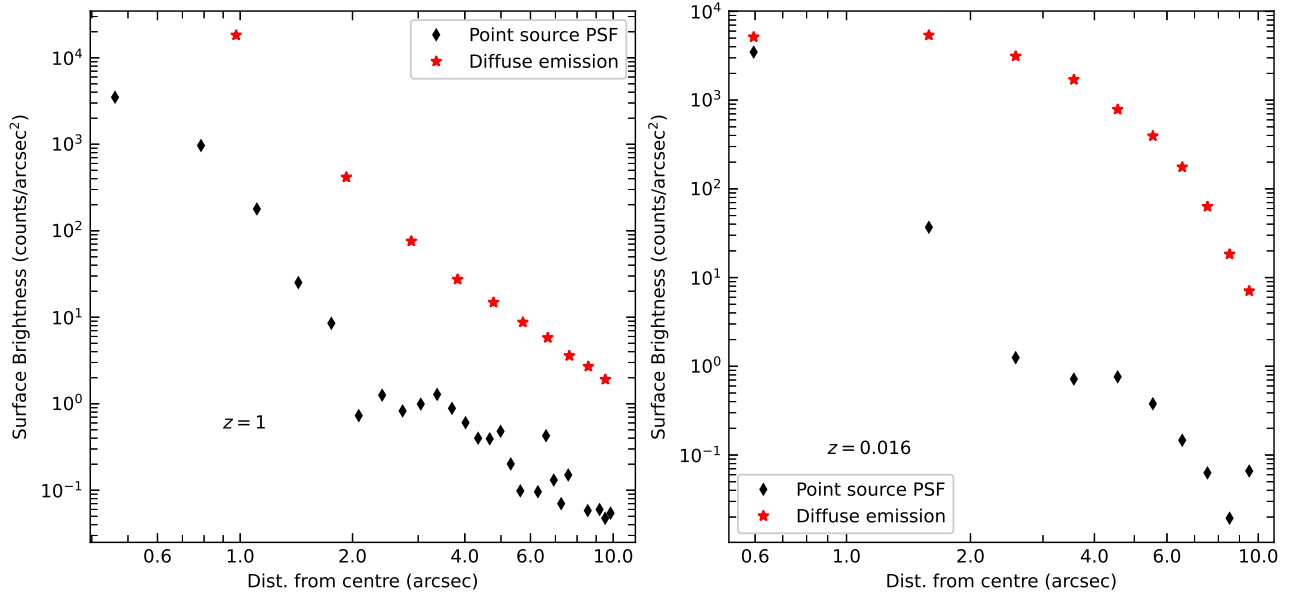}
\caption{Contribution of the central AGN on the diffuse X-ray emission. Left and right panels show the example of one system at $z=1$ and $z=0.016$ respectively. Colour schemes are same in both the panels. Red stars represent the extended X-ray emission from the gas surrounding the AGN where black diamonds represent the X-ray emission coming from the central AGN itself. It can be seen that at the very central pixel AGN contribution is dominating the diffuse X-ray emission. See discussions in Section \ref{sec:3.3}.}
\label{fig:psf}
\end{center}
\end{figure*}

Although we have not considered the X-ray emission coming from the central AGN itself (limitation of the subgrid model), as mentioned earlier, here we try to qualitatively estimate the AGN contribution to assess the detectability of this effect. For that we first convert the bolometric luminosities of the AGN to the corresponding X-ray luminosities using the bolometric correction proposed by \cite{Marconi04}. Then we simulate the point spread function (PSF) using MARX at the central pixel assuming a flat spectrum of the point source at $0.5$ keV. Figure \ref{fig:psf}. shows the point source contribution to the diffuse emission at the central pixel for one system at $z=1$ (left panel) and $z=0.016$ (right panel). It can be seen from this figure that emission from the central AGN significantly suppresses the diffuse emission at the very core ($< 1''$). The central PSF contribution to the diffuse emission has been studied in many previous works \citep[e.g.,][]{Chatterjee15, Mukherjee18}. \cite{Chatterjee15} studied the extended X-ray emission from a sample of normal galaxies and X-ray bright AGN hosts and reported a confusion in detecting the extended X-ray emission due to the extended PSF wings of the central AGN. \citet{Chatterjee15} proposed that a sample of AGN which are not detected in X-rays might be a better sample to study the extended X-ray emission in galaxies and characterize the effect of AGN feedback on this extended emission. In a later work, \cite{Mukherjee18} studied the X-ray surface brightness profiles of the optically selected AGNs in order to minimize the PSF contribution of the X-ray bright AGN, where they found a suppressed X-ray emission in the AGN host galaxies, supporting the results of our work.

\subsection{Reconstruction of Feedback Energy}

Finally, we estimate the maximum energy difference in the presence of different modes of feedback as previously proposed by \cite{Chatterjee15} and discussed in R21. The difference in the surface brightness between two modes can be calculated as 

\begin{equation}
\bigtriangleup I (r) =  I_{i} (r) - I_{wi} (r)
\end{equation}
where $I_{wi}(r)$ is emitted surface brightness at a distance r when no feedback from the AGN is delivered and $I_{i}(r)$ denotes X-ray surface brightness in the presence of different feedback modes at a distance r. Hence, the approximate energy delivered by different modes of feedback can be obtained as follows

\begin{equation}
E = 4 \pi D_{L}^{2} \int 2 \pi r \bigtriangleup I (r) dr
\label{eq:Ef}
\end{equation}
where $D_{L}$ is the luminosity distance.

Using equation \ref{eq:Ef} and Figure \ref{fig:stack_mock}, we calculate the feedback energy for various AGN feedback modes at $z=1$ and $z=0.016$. We find that the integrated feedback energy varies from $1.1 \times 10^{47}$ erg/sec considering all the feedback modes to $\approx 2.8 \times 10^{46}$ erg/sec in case of the ``NoX" and ``NoJet'' mode for the high redshift objects. Value of the integrated feedback energy for the low redshift objects varies a little for different modes of AGN feedback and it is found to be $\approx 10^{41}$ erg/sec.					

By comparing the feedback energy obtained for various modes from our work with those of the actual observation, we can comment on the feedback modes that are at play in the real Universe. Previous studies have been done to measure the feedback energy estimated from the X-ray cavities produced by the AGN \citep{birzan04, Rafferty06, birzan08, Kokotanekov17}. Our results show an agreement with the energy scale at the low redshift obtained from these observations. We find that the suppression of X-ray emission in the surrounding medium of an active galactic nuclei depends on the mode of feedback and the effect is scale dependent. Although it is difficult to observationally distinguish different feedback channels in a particular system, we emphasize that the magnitude of suppression may give us clues toward the mode of feedback that is at action. We further note that cross-correlation studies with other wavelengths (e.g., Sunyaev-Zeldovich signal, see \cite{Chatterjee08}) may also provide stronger constraints on feedback modes of AGN. Thus by comparing the simulated results with the observations it is possible to provide limits on the plausible modes of feedback. This is discussed again in the next section.

\section{Conclusions} \label{sec:conclu}

In this work we have tried to understand the influence of feedback from the AGN on their adjacent gas and the importance of different modes through which the energy outflow from the AGN interacts with their surroundings by analysing the diffuse X-ray emission from galaxy groups and clusters. In addition to this, we examine the robustness of our theoretical model by performing synthetic observations and comparing those with previous results. From the X-ray surface brightness profile of the theoretical model (Figure \ref{fig:stack_th}), we can see a clear deficit in the X-ray emission surrounding the black holes in the presence of all the feedback modes compared to the scenario when no feedback is offered by the central black holes. This result statistically stays consistent for both the high and low redshift objects. Emergence of lower X-ray surface brightness at the centre of the AGN host galaxy groups and clusters is consistent with the works of many previous groups using cosmological simulation as well as actual X-ray observations \citep[e.g.,][]{Gaspari11, Pellegrini12, Mukherjee18, Robson20, RKC21}. \cite{Sun09} studied the gas properties of a sample of 43 galaxy groups at low redshifts ($0.012-0.12$) observed with \textit{Chandra}, among which AGN feedback activity has been detected in some systems from their entropy and surface brightness profiles. Moreover, X-ray cavities resulting from the AGN feedback activity can be detected from the surface brightness profiles of galaxy groups and clusters \citep{HL15, Kokotanekov18, Liu19, Pandge19ApJ, Kolokythas20}.

A major goal of this work is to investigate the importance of different modes of AGN feedback on their surrounding X-ray emitting gas. It has been seen from Figure \ref{fig:stack_th} that inclusion of jet and X-ray mode of feedback with the radiative wind feedback from the AGN is more efficient in reducing the diffuse X-ray emission adjacent to the black holes. The difference in the X-ray surface brightness profile obtained by R21 in their study (Figure 2. of \citealt{RKC21}) in the presence and absence of AGN feedback is small compared to the difference obtained in this work when all the feedback modes are included (red dots) and excluded (green squares) at $z = 1$ (left panel of Figure \ref{fig:stack_th}). We explain this as the consequence of different feedback models used in R21 and SIMBA.

We examine the possibility of detecting AGN feedback at different redshifts using the synthetic observation technique developed by R21. Stacked X-ray surface brightness profiles obtained from the mock \textit{Chandra} observations for different feedback modes at $z = 1$ and $z = 0.016$ are shown in Figure \ref{fig:stack_mock}. It shows no significant difference of the X-ray surface brightness between different feedback modes of high redshift objects within a few central pixels. This can be understood as a consequence of the limited angular resolution of the \textit{Chandra} telescope. However, an excess emission is observed at the outer radii which could be the outcome of the accumulation of the hot gas that is displaced from the black hole neighbourhood by the feedback activity. Detection of this excess emission in the presence of all feedback modes could be a viable method for observing AGN feedback signature in high redshift objects. 

We note that R21 used the D08 simulation to make the statistical analysis of the influence of AGN feedback on their ambient medium, where only the radiative mode of feedback from the AGN is considered in the simulation. On the other hand, SIMBA takes into account jet and X-ray mode along with the radiative mode while modeling the AGN feedback. Thus we see from Figure \ref{fig:stack_th} that radiative wind feedback alone might not be effective in sufficiently evacuating the diffuse X-ray emitting gas around the AGN. Jet and X-ray mode of feedback play substantial roles. Besides, comparing the results of this work with the existing and upcoming X-ray observations it would be possible to comment on the mode of feedback effective in the observed systems. Our results are consistent with the findings of R21 who use the D08 simulation to make similar predictions. We are thus confident about the ubiquity of our results which we observe using multiple simulations.

For the low redshift objects, the right panel of Figure \ref{fig:stack_mock} shows that the difference between the X-ray surface brightness in the presence of different AGN feedback modes can be detected even within the central pixels of the stacked maps. Also, it is observed that X-ray emission decreases significantly at the central regions after incorporating the X-ray feedback, demanding the importance of this feedback mode in the cosmological simulation to properly explain different observables. In a recent work of disentangling the AGN feedback from the stellar feedback, \cite{Chadayammuri22} studied the importance of different modes of feedback on the circumgalactic medium (CGM) using the galaxies from eROSITA Final Equatorial Depth Survey (eFEDS). Their results show a decrement in the X-ray luminosity in the presence of feedback from the AGN, supporting the result of this work. They have also compared their results with different cosmological simulations offering different modes of feedback and commented that kinetic jet feedback needs to be modelled with a dependence on the halo mass.

From this work we conclude that with the specifications of the existing X-ray telescopes, the imprint of AGN feedback reflected from the X-ray surface brightness profile of the diffuse gas inside galaxy groups and clusters, can be resolved for the objects at lower redshift while this signature is unresolved at the central few pixels for the high redshift objects. It is important to discuss the detection possibility of the direct observational signatures of AGN feedback from the X-ray maps of the high redshift objects with the upcoming X-ray missions like \textit{XRISM}, \textit{Athena} as well as the concept mission \textit{Lynx}. \textit{XRISM} and \textit{Athena} are designed to achieve a great field of view, larger effective area and high count rate capabilities \citep{Barcons12, Nandra13, XRISM20, Barret20} that will enhance the detection probabilities of the faint objects such as AGN, galaxy groups and clusters extending to a very high redshift \citep{Cucchetti18, Mernier20, Oppenheimer21, Habouzit22}. However, angular resolution of \textit{XRISM} is $\sim 1.7'$, that is much lower than \textit{Chandra}. While \textit{Athena} is designed to achieve a better resolution of $\sim 5''$, it is still lower than \textit{Chandra}. Hence, it can be qualitatively understood that the signature of AGN feedback from the X-ray surface brightness profile of the high redshift objects can not be resolved with \textit{XRISM} and \textit{Athena} in the vicinity of the black holes. However, \textit{Lynx} is proposed to have an excellent angular resolution of $\sim 0.3''$ \citep{Lynx18} which can enable us to resolve the X-ray surface brightness profiles of the high redshift objects in the presence and absence of the feedback from AGN. We plan to continue the search for detecting the feedback signatures from the high redshift galaxy clusters with the future X-ray telescopes based on the techniques developed in this work and R21 in a future study.

\begin{acknowledgments}

The authors thank the referee for very useful suggestions that helped in significant improvement of the draft. RKC and SC thank Inter-University Centre for Astronomy and Astrophysics (IUCAA) for providing computational support through the Pegasus super computing facility. Part of the computations were performed using research computing facilities offered by Information Technology Services, the University of Hong Kong. This research has made use of the software provided by the Chandra X-ray Center (CXC) in the application package CIAO. RKC and LD acknowledge the support from the Hong Kong Research Grants Council (HKU17305920) and the National Natural Science Foundation of China (HKU12122309). RKC thanks Prof. Romeel Davé and Dr. Dylan Robson for useful discussions on SIMBA. SC acknowledges support from the Department of Science and Technology through the SERB-CRG-2020-002064 grant and from the Department of Atomic Energy for the 57/14/10/2019-BRNS grant.  

\end{acknowledgments}


\bibliography{ref}{}
\bibliographystyle{aasjournal}

\appendix
\counterwithin{figure}{section}

\section{Dependence on Exposure Time}
\begin{figure*}[b]
\begin{center}
\includegraphics[width=17cm]{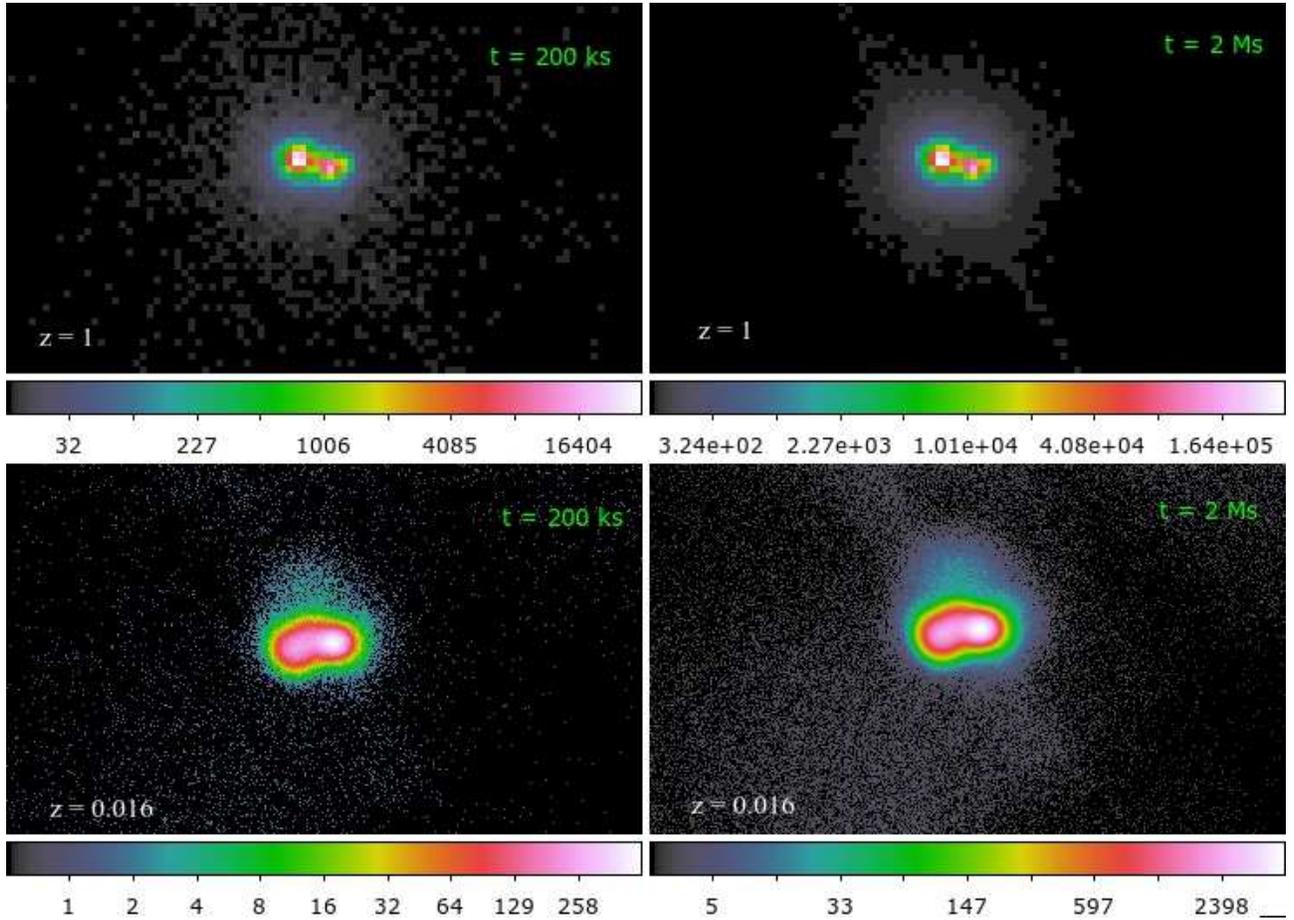}
\caption{X-ray maps for different exposure times. Top left panel: X-ray map of a system at z=1 including all the AGN feedback modes, observed for an exposure time of 200 ksec (same as shown in top left panel of Figure \ref{fig:eg_mock_z1}). Top right panel: X-ray map of the same system observed for 2 Msec. Bottom left panel: X-ray map of a system at z=0.016 including all the modes of feedback from AGN, observed for an exposure time of 200 ksec (same as shown in top left panel of Figure \ref{fig:eg_mock}). Bottom right panel: X-ray map of the same system observed for a higher exposure time of 2 Msec. Colourbars represent photon number counts per pixel in all the panels. An increase of the SNR is observed for the higher exposure times as expected.}
\label{fig:exp}
\end{center}
\end{figure*}

We have performed the synthetic observations of an example system for different exposure times. Figure \ref{fig:exp} represents a system that is observed for 200 ksec (left panel) and 2 Msec (right panel) exposure time both at $z=1$ (top panel) and $z=0.016$ (bottom panel). Here we show the X-ray maps of the system including all the feedback modes at both high and low redshifts observed for two different exposure times. The top and bottom left panels are the same X-ray maps shown in the top left panels of Figure \ref{fig:eg_mock_z1} and Figure \ref{fig:eg_mock} respectively, only shown here with a different colour scheme for visual purpose. Comparing left and right panels, it can be noticed that the signal to noise ratio (SNR) increases moderately with the higher exposure time (right panel) at both redshifts. However, we carry on the analysis throughout this paper using the exposure time of 200 ksec as the SNR is high enough at this exposure time for a significant detection at both $z=1$ and $z=0.016$. 

\end{document}